\documentclass{iopart}
\usepackage{graphicx}
\usepackage{iopams}
\newcommand{\url}{\tt}

\begin{document}

 \letter{Negative Komar Mass of Single Objects in Regular, Asymptotically Flat Spacetimes}
 
 \author{M Ansorg$^1$ and D Petroff$^2$}
 \address{$^1$ Max-Planck-Institut f\"ur Gravitationsphysik, Albert-Einstein-Institut, 14476 Golm, Germany}
 \address{$^2$ Theoretisch-Physikalisches Institut, University of Jena, Max-Wien-Platz 1, 07743 Jena, Germany}
 \eads{\mailto{mans@aei.mpg.de}, \mailto{D.Petroff@tpi.uni-jena.de}}

 \date{\today}

\begin{abstract}
We study two types of axially symmetric, stationary and
asymptotically flat
spacetimes using highly accurate numerical methods. The
one type contains a black hole surrounded by a perfect
fluid ring and the other a rigidly rotating disc of dust
surrounded by such a ring. Both types of spacetime are
regular everywhere (outside of the horizon in the case
of the black hole) and fulfil the requirements of the
positive energy theorem. However, it is shown that both the
black hole and the disc can have negative Komar
mass. Furthermore, there exists a continuous transition
from discs to black holes even when their Komar masses
are negative.
\end{abstract}

\pacs{04.70.Bw, 04.40.-b, 04.25.Dm \hfill preprint number: AEI-2006-057}
\vspace{1cm} 
 
  A great many definitions for mass in General Relativity have been proposed. Many of them
consider only the spacetime as a whole, whereas others assign a mass locally to
a portion of the spacetime. Even the latter are not always applicable to a single
object in a spacetime containing multiple ones.

In \cite{Bardeen73}, Bardeen considers an axially symmetric, stationary spacetime
containing a black hole surrounded by a perfect fluid matter distribution and
assigns to each of the two objects a mass based on the Komar integral \cite{Komar59}.
In a similar way, he assigns an angular momemtum to each of the objects.

Various other local mass definitions include the Hawking mass
\cite{Hawking68}, the Bartnik mass \cite{Bartnik89}, the Christodoulou mass \cite{Christodoulou70},
and a related mass used for isolated horizons, which can be generalized to dynamical ones
\cite{ABL01,AK02}. For a review article see \cite{Szabados04}. A mass definition
applicable both to matter and black holes, and which can be used for single components
of a many-body problem in stationary spacetimes, is the one based on the Komar integral.
We will refer to it here as the Komar mass even when applied to single
objects.
 
In this letter, we study axially symmetric and stationary spacetimes containing
either a black hole or a rigidly rotating disc of dust, each surrounded by a
perfect fluid ring. We find that there exists a continuous transition from the
disc to the black hole. If one is interested in being able to talk about the
mass of the individual objects in such stationary spacetimes and since the
transition from the disc to the black hole exists, then one is led to look
for a definition that is applicable in either scenario. We thus choose to examine
the behaviour of the Komar mass and find that it can become negative both for the
black hole%
\footnote{Negative horizon masses have also been observed for rotating
black holes of Einstein-Maxwell-Chern-Simons theory \cite{KunzNavarro06}.}
and for the disc although the total mass of the spacetime is of course
positive as guaranteed by the positive energy theorem. We find morover
that the continuous transition mentioned above exists even when the Komar
mass of each of the two central objects is negative.

The Poisson equation in Newtonian gravity
\begin{equation}\label{Poisson}
 \nabla^2 U = 4\pi\varepsilon
\end{equation}
relates the potential $U$ to the mass density $\varepsilon$%
\footnote{We use units in which the gravitational constant and
speed of light are equal to one, $G=c=1$.}.
The mass contained in any volume $\mathcal{V}$ of space
can be defined by integrating Eq.~\eref{Poisson} over that
volume and applying the divergence theorem:
\begin{equation}\label{mass_Newtonian}
 M 
  = \int_{\mathcal V}\, \varepsilon\, d^3x  
  = \frac{1}{4\pi}\oint_{\partial \mathcal V} \nabla U \cdot d\boldsymbol{F}.
\end{equation}
Obviously a region of space containing no matter has zero mass
and the total mass can be found using
\[ M_{\rm{tot}} = -\lim_{r \to \infty} (rU).\]
Moreover, mass is additive, i.e.\ the mass contained in a region of space is
simply the sum of the mass contained in each subregion of an arbitrary
subdivision of that space.

A similar procedure can be used to define a relativistic mass
in axially symmetric and stationary spacetimes.
Such a spacetime containing black holes and perfect fluids with
strictly azimuthal motion can be described in
Weyl-Lewis-Papapetrou coordinates by the line element
\[ ds^2 = e^{2\mu}(d\varrho^2 + d\zeta^2) 
       + \varrho^2 B^2 e^{-2\nu}\left(d\varphi - \omega\,dt \right)^2
       - e^{2\nu} dt^2,
\]
where the metric functions depend only on $\varrho$ and $\zeta$.
The energy-momentum tensor of the perfect fluid is
\[ T^{\mu\nu} = (\varepsilon + p)u^\mu u^\nu + p\, g^{\mu\nu},\]
where $\varepsilon$ is the energy density of the fluid, $p$ its pressure and
$u^\mu$ its four-velocity.

The Komar integral for the mass can be constructed by integrating
Einstein's equation
\[ R^t_t = 8\pi \left(T^t_t - \case{1}{2} T\right) \]
over any volume in the 3-space
generated by taking $t=\rm{constant}$. Applying
the divergence theorem again then yields (see \cite{Bardeen73})
\begin{equation}\label{mass_Komar}
M 
  = \int_{\mathcal V} \tilde{\varepsilon}\, d^3x  
  = \frac{1}{4\pi}\oint_{\partial \mathcal V} 
    \left(B\nabla\nu - \frac{1}{2} \varrho^2 B^3 e^{-4\nu}  \omega \nabla \omega  \right)
    \cdot d\boldsymbol{F},
\end{equation}
with
\begin{eqnarray*}
 \tilde{\varepsilon}:=&\, e^{2\mu} B \left[ (\varepsilon+p) \frac{1+v^2}{1-v^2} + 2p
   + 2\varrho B e^{-2\nu}(\varepsilon+p)\,\omega\frac{v}{1-v^2}\right] \\
 v :=&\, \varrho B e^{-2\nu}(\Omega - \omega),
\end{eqnarray*}
where $\Omega$ is the angular velocity of a fluid element with respect to
infinity and the vector operators have the same meaning as in a Euclidean
space in which $(\varrho,\varphi,\zeta)$ are cylindrical coordinates.

The mass defined in the surface integral of Eq.~\eref{mass_Komar} is the Komar mass
(we use the term here even when $\mathcal V$ is of finite extent).
One can see that here too a {\it regular} volume containing no matter
(i.e.\ $\varepsilon=p=0$) has zero mass. If one is considering a black hole spacetime,
then the region interior to the horizon can be excised and the surface integral in
Eq.~\eref{mass_Komar} used to define the mass of the black hole. The mass of any
single object in a stationary  spacetime could thus be defined by calculating the
surface integral in Eq.~\eref{mass_Komar} over a surface containing that object and
only that object. It can be used both for matter and for black holes and has the
additive property familiar from Newtonain theory that the sum of the masses of the single 
objects equals the total (ADM) mass of the spacetime
\[ M_{\rm{tot}} = -\lim_{r \to \infty} (r\nu).\]

A quantity, which plays an important role in black hole thermodynamics
and is constant over the horizon is the surface gravity. In coordinates,
such as the ones chosen in this letter, in which the horizon is a sphere,
it reads
\[ \kappa = e^{-\mu}\frac{\partial}{\partial r}e^{\nu}, 
 \qquad r:=\sqrt{\varrho^2+\zeta^2}.\]
Smarr \cite{Smarr73} showed that for the Komar mass of the black hole
\begin{equation}\label{Smarr}
 M_{\rm h} = \frac{\kappa A}{4\pi} + 2\Omega_{\rm h} J_{\rm h}
\end{equation}
always holds, true even in the presence of a surrounding ring \cite{Bardeen73},
see also \cite{Carter73}.
Here $\Omega_{\rm h}$ is the angular velocity of the black hole (i.e.\ the
constant value of the function $\omega$ over the horizon), $J_{\rm h}$ its
angular momentum and $A$ its area.

A similar expression can be derived for the rigidly rotating disc of dust
(cf.\ III.15 in \cite{BW71}), also valid in the presence of a surrounding ring.
It turns out that the potential $g'_{tt}$ in a coordinate system co-rotating with
the disc (and denoted here by the prime) must be a constant along its `surface'
\begin{equation}\label{gtt}
 -g'_{tt} = e^{2\nu}\left(1-v^2\right) =: e^{2V_0^{\rm d}}.
\end{equation}
This constant is related to the relative redshift $Z_0^{\rm d}$ of photons with zero
angular momentum emitted from the surface of the disc and observed at infinity
via the equation
\[ Z_0^{\rm d} = e^{-V_0^{\rm d}} - 1.\]

Making use of the definition for the baryonic mass
\[M_0 = \int \varepsilon u^t \sqrt{-g}\, d^3x,\]
we can write the disc's Komar mass as
\begin{equation}\label{Smarr_disc}
 M_{\rm d} = e^{V_0^{\rm d}}M_0 + 2\Omega_{\rm d} J_{\rm d}.
\end{equation}

The first term on the right hand side of Eqs~\eref{Smarr} and \eref{Smarr_disc}
is non-negative. In the absence of the ring, the angular velocity and angular
momentum must have the same sign, so that the second term is also non-negative.
If however a ring
is present, it can induce a frame-dragging effect which allows $\Omega$ and
$J$ of the central object to have different signs. If moreover $\kappa$
(or $e^{V_0^{\rm d}}$) becomes sufficiently small, then the expression
for the Komar mass could become negative. Our results
demonstrate that this indeed occurs.

In order to study spacetimes containing a black hole surrounded by a rigidly
rotating ring with constant energy density, we make use of the multi-domain
pseudo-spectral method described in \cite{AP05}. To study the scenario
containing a disc as opposed to a black hole, we
made appropriate modifications to the program.

\begin{figure}
   \hfill \includegraphics{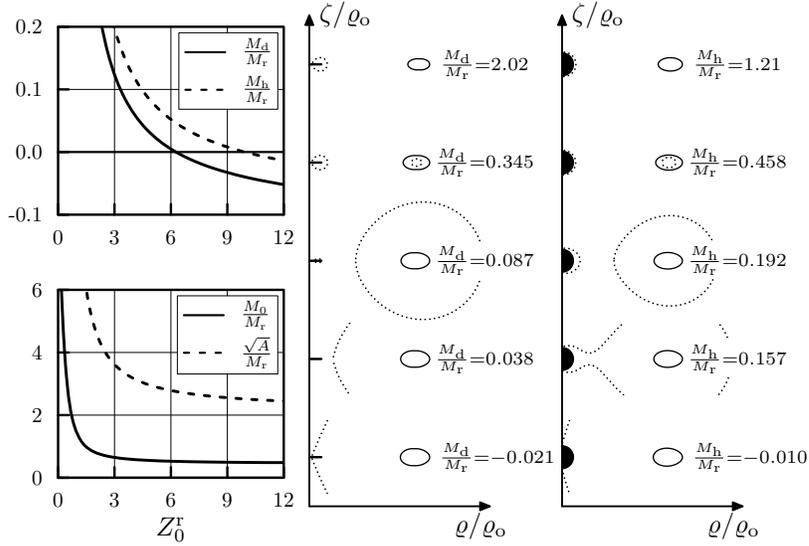}
   \caption{On the upper left,
     the ratio of the Komar mass of the central object to that of the 
     ring is plotted versus $Z_0^{\rm r}$ for a sequence with
     $C_{\rm i}/C_{\rm o}=0.85$, $\beta_o=0.3$ and 
     $\varrho_{{\rm d}}/ \varrho_{\rm o}=0.1$ for the disc or
     $r_{{\rm h}}/ \varrho_{\rm o}=0.1$ for the black hole (see text for an explanation of
     the symbols). Knowing that $M_{\rm r}$ remains positive, one can see that $M_{{\rm d/h}}$
     becomes negative. The lower left shows a similar plot, but containing the disc's baryonic
     mass and the square root of the horizon area. On the right, the coordinate shape of the
     ring and central object (solid lines) and their ergospheres (dotted lines) are drawn for
     these sequences.
     \label{MD_Mc_combined}}
\end{figure}

In the upper plot of Fig.~\ref{MD_Mc_combined} we consider sequences of configurations that
demonstrate the existence of negative Komar masses for discs and black holes
surrounded by rings. In the absence of a ring, the analytic solution for the
disc is known \cite{NM95} and depends on one `physical' and one scaling parameter
as does the Kerr metric. When a ring is present, the
configurations are described by four physical parameters, so that a sequence can be
specified by holding three parameters constant and varying a fourth.
Along the sequences in the plot, the ratio of proper inner to outer circumference of the
ring was held at a value of $C_{\rm i}/C_{\rm o}=0.85$ and a mass-shed parameter%
\footnote{The definition of $\beta_{\rm o}$ is 
           $\frac{2\varrho_{\rm o}}{\varrho_{\rm o} -\varrho_{\rm i}}
             \left. \frac{d(\zeta_{\rm B}^2)}{d(\varrho^2)}
             \right|_{\varrho=\varrho_{\rm o}}$ where
           $\zeta_{\rm B} = \zeta_{\rm B}(\varrho)$ is a parametric
           representation of the surface of the ring. The value
           $\beta_{\rm o}=0$ corresponds to the outer mass-shedding limit.
}
for the outer edge of the ring was held at a value $\beta_o=0.3$. For the
disc sequence, the ratio of its coordinate radius to that of the outer edge
of the ring was chosen to be $\varrho_{\rm d}/\varrho_{\rm o}=0.1$, whereas for
the black hole sequence $r_{\rm h}/\varrho_{\rm o}=0.1$ was chosen. This last
parameter choice, which refers to the radius of the black hole $r_{\rm h}$, is possible
since coordinates have been chosen in which the black hole is always a sphere.
In the figure, various quantities are plotted versus the relative redshift $Z_0^{\rm r}$
of the ring, which is defined as in Eq.~\eref{gtt}. This quantity is also discussed in
\cite{AKM4}, in which relativistic rings without a central object are examined.
The figure shows that
the ratio of the central mass to the ring mass $M_{{\rm d/h}}/M_{\rm r}$ indeed becomes negative
($M_{\rm r}$ remains positive throughout). It is important to emphasize that
$M_{\rm{ d/h}}=0$ in this plot does not correspond to a vanishing central object. 
This fact is demonstrated in the lower plot of Fig.~\ref{MD_Mc_combined} which shows that
neither the disc's baryonic mass $M_0$ nor the black hole's horizon area $A$ tends to
zero when compared to the ring's Komar mass for large $Z_0^{\rm r}$.

The evolution of the coordinate shape of the ring, the central object and their ergospheres
can be followed on the right of Fig.~\ref{MD_Mc_combined}. Looking first at the series
of pictures on the left, we begin with a fairly
`Newtonian' ring (i.e.\ with small $Z_0^{\rm r}$) and can see that only the disc
possesses an ergosphere. The ring and the disc are counter-rotating, meaning their
angular velocities have opposite signs. As the ring becomes increasingly relativistic
and develops an ergosphere, its frame-dragging effect on the disc becomes more pronounced,
causing the disc's angular velocity to decrease, whence its ergosphere shrinks and
finally vanishes. As $Z_0^{\rm r}$ increases further, the frame-dragging finally forces the
disc to co-rotate with the ring although its angular momentum still has the opposite
sign. Relative to the size of the ring, the
ergosphere grows very large (hence we show only a portion of its boundary
in the last two pictures in the sequence).
From an outside observer's perspective, the configuration
is shrinking toward the centre and the outside metric beginning to resemble that
of the extreme Kerr metric.
The ring's ergosphere continues to
grow, finally engulfing the disc. After a good portion of the disc finds itself
inside the ergosphere, the frame dragging becomes significant enough that the
magnitude of $\Omega_{\rm d}\, J_{\rm d}$ is
sufficiently large to result in a negative mass.
The series of pictures on the right is the counterpart 
for a black hole and shows similar behaviour to the disc case. A black hole with non-vanishing
$\Omega_{\rm h}$ always has an ergosphere surrounding it however. Its sense of rotation
must agree with that of the ring before their ergospheres merge, independent
of the sign of $J_{\rm h}$. 

\begin{figure}
   \hfill \includegraphics{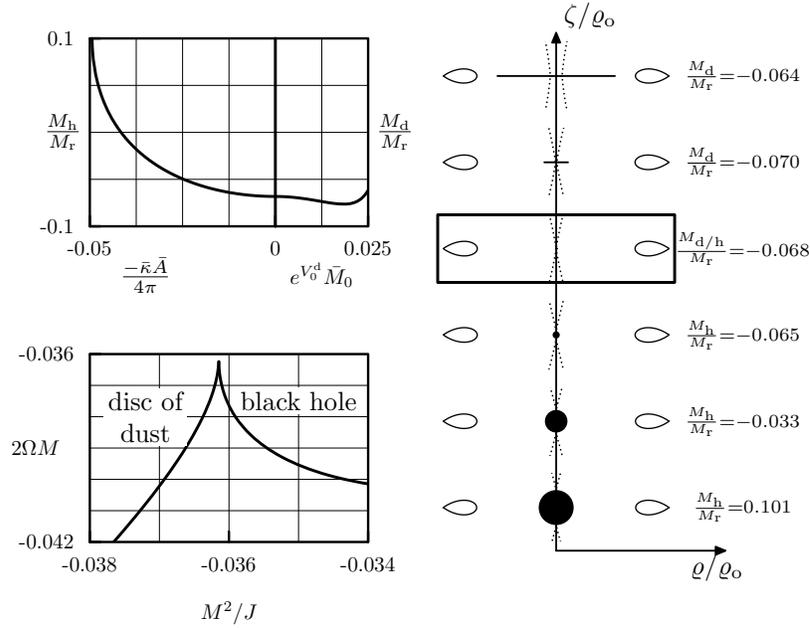}
   \caption{On the upper left, the ratio of the Komar mass of the central object
     to that of the ring is plotted versus a measure of the distance to the degenerate
     black hole solution. On the lower left, a plot similar to Fig.~2 in \cite{NM93}
     is shown for the transition. On the right, the coordinate shape of the ring and central
     object (solid lines) and their ergospheres (dotted lines) are drawn.
     The framed picture indicates the transition point from the disc to the
     black hole.\label{transition_combined}}
\end{figure}

The parametric transition from a rigidly rotating disc (without a surrounding ring)
to a black hole has been studied numerically \cite{BW71} and analytically \cite{Meinel02}.
A generalization of the proofs in \cite{Meinel04, Meinel06}
implies that such a transition also exists for a disc surrounded by a ring
if and only if $V_0^{\rm d}$ of the disc tends to $-\infty$, which in turn implies that
$M_{\rm d} = 2\Omega_{\rm d} J_{\rm d}$ must hold%
\footnote{The generalization of the arguments in \cite{Meinel06} requires the
assumption that $-\xi^i u_i$ as defined there is bounded from below. In the
presence of a surrounding ring, $\eta^i u_i$ need not have the same sign as
$\Omega_{\rm d}$.}.
The equality of Eqs~\eref{Smarr_disc} and
\eref{Smarr} then requires for non-vanishing black holes that $\kappa=0$.
The upper plot on the left of Fig.~\ref{transition_combined} shows that
such transitions do indeed exist. Here the mass ratio was chosen as an exemplary
parameter and plotted versus a measure of the distance to the transition point
representing a degenerate black hole surrounded by a ring. Both sequences are
defined by $\beta_o=0$, $V_0^{\rm r}=-2.7$ and $C_{\rm i}/C_{\rm o}=0.85$.
A bar over a quantity indicates that it has been made dimensionless through
multiplication with the appropriate power of the ring's density $\varepsilon$.
The lower plot suggests a very similar
transition to the one known analytically for the rigidly rotating disc of dust without
a ring as can be seen by comparing it to Fig.~2 in \cite{NM93}, in which an
interpretation in terms of a phase transition was considered.
The picture sequence on the right of Fig.~\ref{transition_combined}
shows the evolution of the coordinate shapes of these configurations.

References to `the mass' of a single constituent of a many-body
system are accepted by convention in some branches of General Relativity,
such as within the binary black hole community. In stationary spacetimes,
one may also wish to be able to refer to individual masses and, indeed,
the Komar mass can be defined rigorously and has various attractive features,
perhaps the most important one being that it can
be used on either side of the transition from matter to a black hole. The fact
that it can become negative is related to the fact that the definition involves
both local quantities and a reference through $\Omega$ to asymptotic infinity.
It thus seems unlikely that locally unusual properties will be observed, but
an investigation of geodesic motion in the vicinity of such objects would be
interesting. Other interesting questions such as the minimal attainable mass
ratio (i.e.\ how close it can come to $-1$) will be the topic of later work.
Finally, we want to point out that various authors (e.g.\ \cite{RMRS05})
have considered negative Komar masses to be unphysical and we hope that the
present work shows that this need not be the case.

\ack
   We are very grateful to R.~Meinel, A.~Ashtekar, S.~Bonazzola, B.~Carter, J.~Ehlers,
   E.~Gourgoulhon, J.~Jaramillo, G.~Neugebauer and L.~Rezzolla for fruitful discussions.
   This work was supported in part by the Deutsche Forschungsgemeinschaft (DFG)
   through the SFB/TR7 ``Gravitationswellenastronomie''. 
  
\section*{References}

 \bibliographystyle{unsrt}
 \bibliography{Reflink}
  
\end{document}